\documentclass[10pt,conference]{IEEEtran}

\usepackage{ifpdf}

\usepackage{cite}

\ifCLASSINFOpdf
  \usepackage[pdftex,hiresbb]{graphicx}
\else
\fi

\usepackage[cmex10]{amsmath}
\usepackage{fixltx2e}

\usepackage{url}

\usepackage{enumitem}
\usepackage{multirow}
\usepackage{bigstrut}
\usepackage{amssymb,amsfonts}
\usepackage{bm}
\usepackage[pdftex]{xcolor}
\usepackage[normalem]{ulem}
\usepackage{xspace}
\usepackage{textcomp}
\usepackage{algorithm}
\usepackage{algpseudocode}
\algrenewcommand\alglinenumber[1]{\footnotesize #1:}
\usepackage{xpatch}
\usepackage[caption=false]{subfig}
\usepackage{placeins}

\makeatletter
\newcommand{\removelatexerror}{\let\@latex@error\@gobble}
\makeatother

\newcommand{\rbt}{\mathbf{r}}
\newcommand{\ubt}{\mathbf{u}}
\newcommand{\wbt}{\mathbf{w}}
\newcommand{\ybt}{\mathbf{y}}
\newcommand{\Hbt}{\mathbf{H}}
\newcommand{\Ibt}{\mathbf{I}}
\newcommand{\xbt}{\mathbf{x}}
\newcommand{\Xbt}{\mathbf{X}}
\newcommand{\zero}{\mathbf{0}}

\newcommand{\changecolor}[1]{{\color{black}#1}}

\makeatletter
\xpatchcmd{\algorithmic}{\itemsep\z@}{\itemsep=0.6ex plus 0.6ex  minus 0.6ex}{}{}
\makeatother

\makeatletter
\def\paragraph{\@startsection{paragraph}{4}{\z@}{0.2ex plus 0.2ex minus 0.2ex}%
{0ex}{\normalfont\small\bfseries}}
\makeatother

\IEEEoverridecommandlockouts\IEEEpubid{\makebox[\columnwidth]{ 979-8-3503-1090-0/23/\$31.00~\copyright2023 IEEE \hfill} \hspace{\columnsep}\makebox[\columnwidth]{ }}

\makeatletter
\newcommand\fs@betterruled{%
  \def\@fs@cfont{\bfseries}\let\@fs@capt\floatc@ruled
  \def\@fs@pre{\vspace*{5pt}\hrule height.8pt depth0pt \kern2pt}%
  \def\@fs@post{\kern2pt\hrule\relax}%
  \def\@fs@mid{\kern2pt\hrule\kern2pt}%
  \let\@fs@iftopcapt\iftrue}
\floatstyle{betterruled}
\restylefloat{algorithm}
\makeatother

\begin{document}
\title{Near-optimal stochastic MIMO signal detection with a mixture of $t$-distributions prior}

\author{%
\IEEEauthorblockN{Junichiro Hagiwara\IEEEauthorrefmark{2}, Kazushi Matsumura, Hiroki Asumi, Yukiko Kasuga, Toshihiko Nishimura, \\ Takanori Sato, Yasutaka Ogawa, and Takeo Ohgane}
\IEEEauthorblockA{%
Graduate School / Faculty of Information Science and Technology, Hokkaido University \\
Kita 14, Nishi 9, Kita-ku, Sapporo, Hokkaido, 060-0814 Japan \\
Email: \{jhagiwara, kazu4.matsumura, asumi, kasuga, nishim, tksato, ogawa, ohgane\}@m-icl.ist.hokudai.ac.jp
}
\IEEEauthorblockA{%
\IEEEauthorrefmark{2}\changecolor{Current affiliation: Faculty of Social Informatics, Mukogawa Women's University}\\
}
}

\maketitle

\begin{abstract}
Multiple-input multiple-output (MIMO) systems will play a crucial role in future wireless communication, but improving their signal detection performance to increase transmission efficiency remains a challenge. 
To address this issue, we propose extending the discrete signal detection problem in MIMO systems to a continuous one and applying the Hamiltonian Monte Carlo method, an efficient Markov chain Monte Carlo algorithm. 
In our previous studies, we have used a mixture of normal distributions for the prior distribution. In this study, we propose using a mixture of $\bm{t}$-distributions, which further improves detection performance. 
Based on our theoretical analysis and computer simulations, the proposed method can achieve near-optimal signal detection with polynomial computational complexity. This high-performance and practical MIMO signal detection could contribute to the development of the 6th-generation mobile network.
\end{abstract}


\marginpar{\hspace{-0.5cm}\rotatebox{90}{{\sffamily\scriptsize GLOBECOM 2023 - 2023 IEEE Global Communications Conference \textbar~979-8-3503-1090-0/23/\$31.00~\copyright2023 IEEE \textbar~DOI: 10.1109/GLOBECOM54140.2023.10437162}}}
\section{Introduction} \label{sec:Introduction}
Wireless communication has become a crucial component of modern society's infrastructure, as evidenced by the widespread use and growth of 5G mobile network. To further improve convenience, research and development of 6G is underway to effectively use radio resources. Among the various technologies that play \changecolor{an} essential role in the effective use of radio resources, multiple-input multiple-output (MIMO) technology is expected to make a significant contribution. The MIMO technology enhances wireless transmission efficiency using multiple antennas at both transmitting and receiving stations.
Although MIMO with a relatively small number of antennas is already in general use, research on massive MIMO with numerous antennas just started receiving attention~\cite{An_Overview_of_Massive_MIMO}.

In MIMO systems, accurately detecting received signals is critical for improving transmission efficiency. Maximum likelihood decoding, which can achieve ideal performance, is computationally impractical for detecting received signals due to its examination of all possible combinations of transmitted signals~\cite{chockalingam2014large}. Meanwhile, linear detection techniques, e.g., the minimum mean-square error method, are computationally less expensive but underperform~\cite{chockalingam2014large}.
The challenge lies in improving the signal detection performance while maintaining realistic computational complexity requirements. One solution is to approach signal detection from a stochastic perspective. This allows for the theoretical consideration of uncertainty in the problem and the application of state-of-the-art probabilistic and statistical techniques.

The problem of MIMO signal detection has been approached from a stochastic perspective using several techniques, such as Bayesian methods~\cite{Fifty_Years_of_MIMO_Detection,Massive_MIMO_Detection_Techniques}. However, to the best of our knowledge, continuous prior distribution has been regarded as inappropriate for MIMO signal detection, except for our previous proposals~\cite{WPMC,AsumiRCS}.
The mixed Gibbs sampling (MGS) method~\cite{MGS} approximates the posterior distribution using a large number of samples, where a discrete distribution is set as the prior distribution and the algorithm is based on Gibbs sampling~\cite{Gibbs}, a type of Markov chain Monte Carlo (MCMC).
The expectation propagation (EP) method~\cite{EP} also sets a discrete distribution as the prior distribution and uses the EP algorithm~\cite{EPMinka} to estimate the parameters of the approximate posterior distribution.
Moreover, the prior distribution in~\cite{VB} is set as a mixture of truncated normal distributions.
This method uses a variational Bayesian algorithm to determine the parameters of the approximate distribution for the posterior distribution.
Previous studies on MIMO signal detection that assume stochasticity have set a discrete or discontinuous distribution as the prior distribution based on the assumption of discrete transmission symbols. 

We extend the discrete signal detection problem in MIMO systems to a continuous one and apply the Hamiltonian Monte Carlo (HMC) method~\cite{DUANE1987216}, an efficient MCMC algorithm for continuous problems.
In our previous studies~\cite{WPMC,AsumiRCS}, we set a mixture of normal distributions as the prior distribution. In this study, we propose a novel method using a mixture of $t$-distributions, which further improves signal detection performance.
Through our theoretical analysis and computer simulations, the proposed method is shown to achieve near-optimal signal detection with polynomial computational complexity.
Our novel MIMO signal detection method will contribute to both the practical and theoretical aspects of future wireless communications.

The remainder of this article is structured as follows:
Section~\ref{sec:Problem formulation} presents a stochastic formulation of the problem.
Section~\ref{sec:Previous work} briefly describes the MGS and EP methods as existing baseline methods.
Section~\ref{sec:Stochastic signal detection with a mixture of $t$-distributions prior} explains the details of the proposed method.
Section~\ref{sec:Numerical results and discussion} discusses our theoretical analysis and computer simulation results.
Finally, we present a summary in Section~\ref{sec:Conclusions}.

The following notations are used in this study.
$\mathbb{C}$ denotes the field of complex numbers.
$\mathrm{Re}(\xbt)$ and $\mathrm{Im}(\xbt)$ denote the real and imaginary parts of $\xbt$, respectively.
$\hat{\xbt}$ denotes the estimate of $\xbt$.
$\lfloor x \rfloor$ denotes the largest integer less than or equal to $x$.
$\zero$ and $\Ibt$ denote the zero vector and unit matrix, respectively.
$\Xbt^\mathrm{T}$ denotes the transpose of $\Xbt$.
The symbol $\sim$ indicates that the random variable on its left-hand side follows the probability distribution on its right-hand side.
The symbol $\propto$ represents a proportional relationship.

\section{Problem formulation} \label{sec:Problem formulation}
\subsection{System model}
Suppose a full-stream transmission in a MIMO system with $N$ transmission antennas and $M$ receiving antennas.
The following relationship holds:
\begin{equation}
\qquad\qquad\qquad\qquad \ybt = \Hbt \ubt + \wbt, \quad \wbt \sim \mathcal{CN}(\zero, \sigma_w^2 \Ibt),
\label{eq:system}
\end{equation}
where the receiving antenna's symbol vector is $\ybt = [y_1, \ldots, y_M]^\mathrm{T} \in \mathbb{C}^M$, the channel matrix is $\Hbt \in \mathbb{C}^{M \times N}$, the transmission antenna's symbol vector is $\ubt = [u_1, \ldots, u_N]^\mathrm{T} \in \mathbb{C}^N$, the noise vector is $\wbt = [w_1,  \allowbreak \ldots, w_M]^\mathrm{T} \in \mathbb{C}^M$, the noise variance is $\sigma_w^2$, and $\mathcal{CN}$ represents a circularly symmetric complex normal distribution.
For manipulation ease, we split complex numbers into their real and imaginary components as follows:\rule[-1.7ex]{0ex}{0ex}
$\ybt \rightarrow \left[ \begin{smallmatrix} \mathrm{Re}(\ybt) \\ \mathrm{Im}(\ybt) \end{smallmatrix}\right]$,
$\Hbt \rightarrow \left[ \begin{smallmatrix} \mathrm{Re}(\Hbt) & -\mathrm{Im}(\Hbt)\\ \mathrm{Im}(\Hbt) & \phantom{-}\mathrm{Re}(\Hbt) \end{smallmatrix}\right]$,
$\ubt \rightarrow \left[ \begin{smallmatrix} \mathrm{Re}(\ubt) \\ \mathrm{Im}(\ubt) \end{smallmatrix} \right]$, and
$\wbt \rightarrow \left[ \begin{smallmatrix} \mathrm{Re}(\wbt) \\ \mathrm{Im}(\wbt) \end{smallmatrix} \right]$.\rule[-2ex]{0ex}{0ex} 
As a supplementary note, the symbols for vectors and matrices remain unchanged hereafter when $N$ and $M$ are reconsidered as $2N$ and $2M$, respectively. In this study, the channel matrix $\Hbt$, noise variance $\sigma_w$, and received signal $\ybt$ are assumed to be known. The transmitted symbols are also assumed to be uniformly random across antennas due to the use of scramblers. Signal detection entails estimating the transmitted symbol vector $\ubt$, given the received signal $\ybt$.

By introducing a stochastic interpretation, ``posterior distribution $\propto$ likelihood $\times$ prior distribution'' holds according to Bayes' theorem:
\begin{equation}
p(\ubt \mid \ybt) \propto p(\ybt \mid \ubt) p(\ubt).
\label{eq:Bayse}
\end{equation}
The goal of stochastic signal detection is to obtain a point estimate of the posterior distribution $p(\ubt \mid \ybt)$.

\subsection{Likelihood}
Equation~\eqref{eq:system} shows $p(\ybt \mid \ubt) = \mathcal{N}(\ybt; \Hbt \ubt, \sigma_w^2 \Ibt)$, where $\mathcal{N}$ represents a real-value normal distribution density.

\subsection{Prior distribution}
\begin{figure}[!t]
\centering
\includegraphics[clip,width=0.8\linewidth]{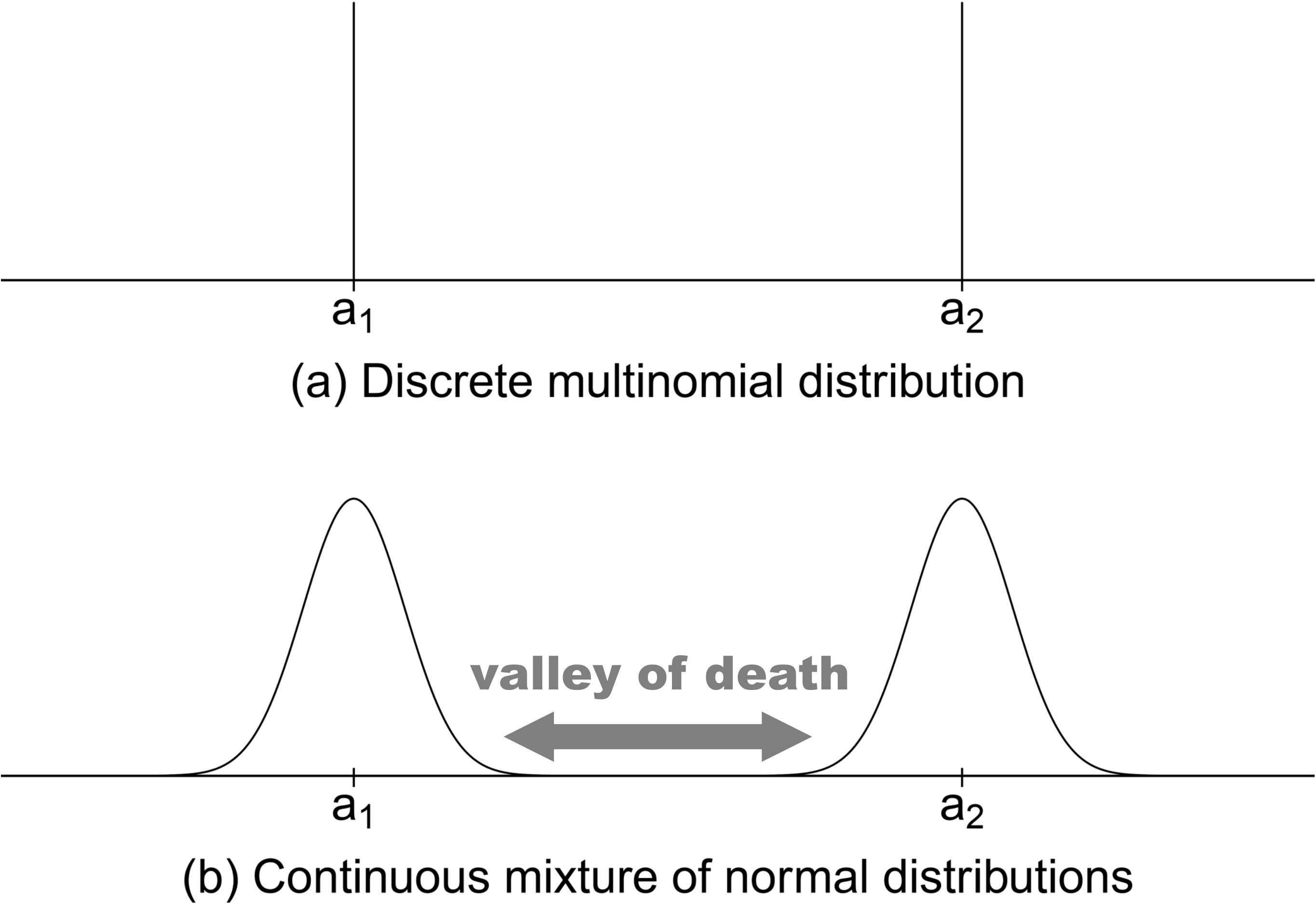}
\caption{Two examples of priors for BPSK.}
\label{fig:prior}
\end{figure}

Equation~\eqref{eq:Bayse} implies that the prior distribution corresponds to regularization terms for correcting the likelihood.
In signal detection, the prior distribution is typically a discrete multinomial distribution that reflects the possibility for discrete signal points. This improves the accuracy of the posterior distribution's estimation by considering the priority at the transmission signal point, as in \eqref{eq:multinomial} (see also Fig.~\ref{fig:prior} (a)):
\begin{equation}
p(\ubt) =  \prod^{2N}_{n=1} \frac{1}{q}\{\delta(u_n - a_1) + \cdots +\delta(u_n - a_q) \},
\label{eq:multinomial}
\end{equation}
where $q$ denotes the square root of the modulation order, $a_1, \ldots, a_q$ denotes the real-valued coordinate of the transmission signal points, and $\delta(x)$ denotes the unit probability mass at $x$.
For example, the MGS and EP methods assume such a prior distribution.

Assuming a continuous distribution for the prior distribution transforms the above problem into a continuous value problem. For example, the following equation demonstrates the application of a mixture of normal distributions (refer to Fig.~\ref{fig:prior} (b)):
\begin{equation}
p(\ubt) = \prod^{2N}_{n=1}  \frac{1}{q}\{\mathcal{N}(u_n; a_1, \sigma) + \cdots + \mathcal{N}(u_n; a_q, \sigma) \},
\label{eq:mixturenormal}
\end{equation}
where $\sigma^2$ denotes a variance of a component's normal distribution and tuning matter.
In our previous studies~\cite{WPMC,AsumiRCS}, we used a mixture of normal distributions as the prior distribution and set the optimal value of $\sigma$ to minimize the bit error rate (BER) through a preliminary search. The optimal value of $\sigma$ corresponds to the equilibrium point where the search efficiency is the best. If $\sigma$ is too large, the search efficiency is low as areas other than the transmission signal points are unnecessarily explored. Meanwhile, if $\sigma$ is too small, there is little overlap in the component's distribution, making it difficult to explore other possible transmission signal points (the ``valley of death'' in Fig.~\ref{fig:prior} (b)).

\subsection{Posterior distribution}
When the prior distribution is assumed to be a mixture of any distribution, it is impossible to obtain the posterior distribution analytically in closed form~\cite{bishop:2006:PRML}. Regardless of whether the component distribution is discrete or continuous, a numerical approximation algorithm must be used to derive the posterior distribution. The point estimate $\hat{\ubt}$ of the posterior distribution can produce multiple candidates, depending on the approximation algorithm used. In addition, the $\hat{\ubt}$ may differ from the original transmission signal point. For instance, when using a continuous distribution as the prior distribution, the search may include areas other than discrete transmission signal points, leading to deviations. To address this issue, we compute the likelihood $p(\ybt \mid \tilde{\ubt})$ after quantizing $\hat{\ubt}$ to the nearest transmission signal point $\tilde{\ubt}$ and consider $\tilde{\ubt}$ with the highest likelihood as the final point estimate.

\section{Previous work} \label{sec:Previous work}
\subsection{MGS method~\cite{MGS}}
The MGS method approximates the posterior distribution by generating a large number of samples. 
It employs Gibbs sampling to search more intensively in regions with higher posterior probability densities.
The MGS method improves the search efficiency by mixing the initialization of the search values with a probability of $1/(2N)$, which solves the problem of getting stuck in local optima.
This approach can be seen as a virtual parallelization of the Markov chains.
The multiple restarts technique, also proposed in~\cite{MGS}, involves running multiple MGS methods with different initial values for the Markov chain and selecting the result with the highest likelihood.
It was shown that using a sufficient number of restarts can result in near-optimal performance.

The computational complexity of the MGS method is $\mathcal{O}(MN)$ per step of the Markov chain, as the computation of the likelihood is the main factor that contributes to its complexity.
With $L_\text{MGS}$ as the total steps in the Markov chain, the final computational complexity is $\mathcal{O}(L_\text{MGS} MN)$.

\subsection{EP method~\cite{EP}}
The EP method approximates the posterior distribution using an uncorrelated multivariate normal distribution $q(\ubt)$.
In this method, the EP algorithm is used to find the parameters that minimize the Kullback--Leibler divergence $- \int p(\ubt \mid \ybt)  \ln\{\allowbreak q(\ubt)/p(\ubt \mid \ybt)\} \mathrm{d}\ubt$.
Specifically, the mean and variance parameters are iteratively refined, and the final mean parameter is used to estimate the transmitted symbols. 
According to~\cite{EP}, the total number of iterations, $L_\text{EP}$, can be set to at most $L_\text{EP}=10$ to reach the maximum detection performance.

The computational complexity of the EP method is $\mathcal{O}(N^3)$ per iteration, as the inverse matrix operation of $\Hbt$ is the main factor that contributes to its complexity. 
With a total of $L_\text{EP} = 10$ iterations assumed, the final computational complexity becomes $\mathcal{O}(10 N^3)$.

\section{Stochastic signal detection with a mixture of $t$-distributions prior} \label{sec:Stochastic signal detection with a mixture of $t$-distributions prior}
\subsection{Setting the prior distribution}
\begin{figure}[!t]
\centering
\includegraphics[width=0.8\linewidth]{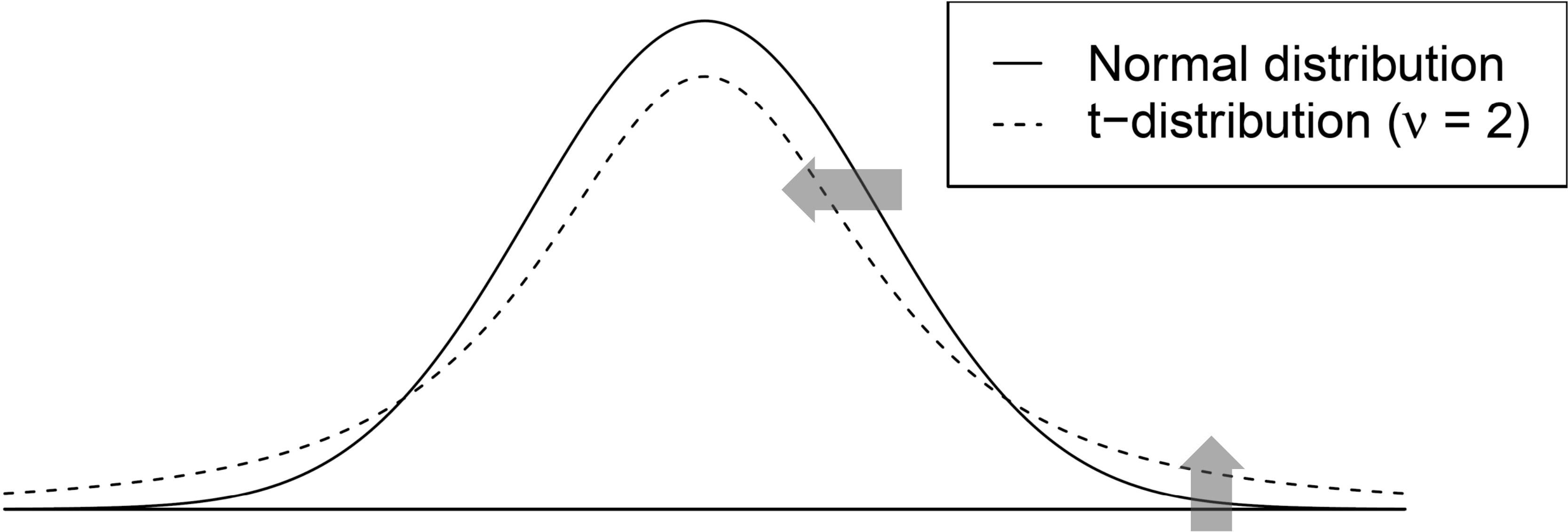}
\caption{The normal and $t$-distributions with the same scale parameter $\sigma$.}
\label{fig:t}
\end{figure}

In this study, we propose the use of a mixture of $t$-distributions as the prior distribution in signal detection as in \eqref{eq:mixturet}:
\begin{equation}
p(\ubt) = \prod^{2N}_{n=1} \frac{1}{q} \{\mathcal{T}(u_n; a_1, \sigma, \nu) + \cdots + \mathcal{T}(u_n; a_q, \sigma, \nu) \},
\label{eq:mixturet}
\end{equation}
where $\mathcal{T}$ denotes a real-valued $t$-distribution density~\cite{bishop:2006:PRML}. The scale parameter $\sigma$ and the degrees of freedom $\nu$ are adjustable or \changecolor{tunable} parameters.
Regarding these, we employ the optimal values obtained through preliminary searches.
The $t$-distribution has a narrower peak and thicker tail than the normal distribution with the same scale parameter $\sigma$ (Fig.~\ref{fig:t}).
Therefore, compared with the normal distribution, the $t$-distribution can more thoroughly search around the transmission signal points.
It also actively explores other potential transmission signal points by overcoming the ``valley of death'' between them. 
As a result, it is expected to have exploration properties superior to our previously proposed mixture of normal distributions~\cite{WPMC,AsumiRCS}.

\changecolor{%
Furthermore, the $t$-distribution is suitable for discrete signal detection from the perspective of sparse estimation. 
The prior distribution in Bayesian estimation corresponds to the constraint (i.e., regularization term) in sparse estimation.
For example, the sparse regression method known as Lasso, proposed in~\cite{LASSO}, applies the $\ell_1$ norm to the regularization term in estimating regression coefficients that are mostly zero but rarely take finite values.
This regularization term with the $\ell_1$ norm corresponds to setting the Laplace (double exponential) distribution centered at zero as a prior distribution (Fig.~\ref{fig:Laplace}).
\begin{figure}[!t]
\centering
\includegraphics[width=0.8\linewidth]{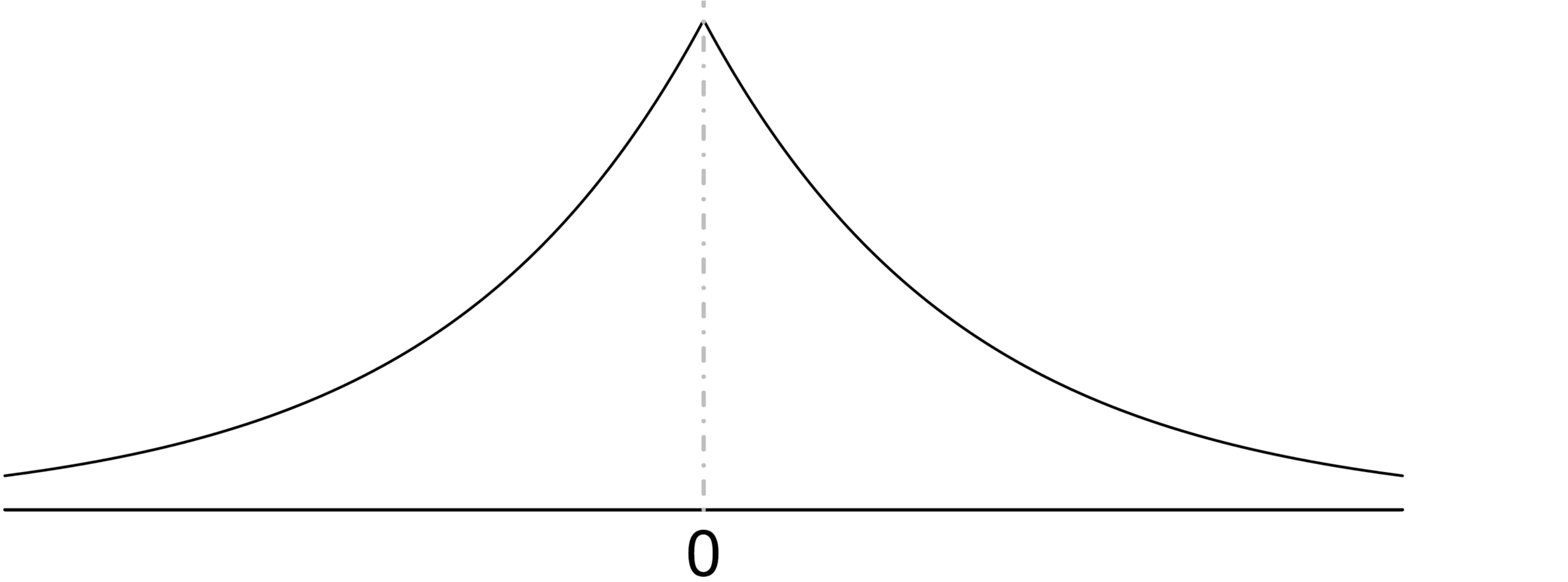}
\caption{\changecolor{Laplace (double exponential) distribution.}}
\label{fig:Laplace}
\end{figure}
The kernel part of the standard Laplace distribution can be expressed as follows:
\begin{align}
& \mathrel{\phantom{=}} \exp \{-|x|\} = \exp \{-\max (-x, x)\} \notag \\
& \sim \exp \{-\text{smooth\_max}(-x, x)\} = \exp \{-\text{LogSumExp}(-x, x)\}\notag \\
& = \exp \left\{-\log \left(e^{-x}+e^x\right)\right\} =\left(e^{-x}+e^x\right)^{-1} \propto \{\operatorname{cosh}(x)\}^{-1} \notag \\
& \sim \textstyle (1+\frac{x^2}{2 !})^{-1}.
\end{align}
As the kernel part of the standard $t$-distribution is $(1+\frac{x^2}{\nu})^{-(\nu+1)/2}$, the $t$-distribution can approximate the Laplace distribution well with $\nu \sim 2$.
Therefore, setting the $t$-distribution centered at the signal point as a prior distribution can be interpreted as allowing sparse estimation based on the $\ell_1$ norm around the signal point.
}

\subsection{Approximation algorithm for the posterior distribution}
In signal detection, setting a continuous distribution as the prior distribution yields a continuous posterior distribution, as the likelihood is a continuous normal distribution. Thus, effective approximation algorithms for continuous problems can be used in estimating the posterior distribution. In our previous study~\cite{AsumiRCS}, we compared the results of using Newton's method, automatic differentiation variational inference method~\cite{ADVI}, and the HMC method as the approximation algorithm, under the condition of a mixture of normal distributions as the prior distribution. The results showed that the HMC method mostly achieved superior signal detection performance with the same computational complexity. This is because the HMC algorithm randomly initializes the searching value at each step of the Markov chain, which helps avoid local optima. Thus, we employ the HMC method as the approximation algorithm for the posterior distribution in this study.

The HMC method, being a type of MCMC, uses the Markov chain mechanism like the MGS method. 
It explores areas with higher posterior probability densities to generate samples that approximate the posterior distribution. Compared with other MCMC methods, the HMC method is efficient in terms of sampling due to its innovative use of the Hamiltonian mechanics framework.
The HMC method intentionally adds a quantity, $\rbt = \mathrm{d}\ubt/\mathrm{d}\tau$ (where $\tau$ is a virtual time), that corresponds to the momentum of $\ubt$ in addition to the variable $\ubt$ being estimated. 
A summary of the HMC method is provided below. 
In this method, the system's potential energy, $U$, and kinetic energy, $K$, are defined as $U(\ubt) = -\ln(p(\ubt \mid \ybt))$ and $K(\rbt) = 1/2 ||\rbt||^2$, respectively. 
The Hamiltonian is introduced as follows:
\begin{equation}
H(\ubt, \rbt) = U(\ubt) + K(\rbt),
\label{eq:Hamiltonian}
\end{equation}
which represents the total energy of the system.
Then, Hamilton's equations are expressed using these two partial differential equations:
\begin{equation}
\begin{split}
\frac{\mathrm{d}\ubt}{\mathrm{d}\tau} & =  \frac{\partial H(\ubt, \rbt)}{\partial\rbt} = \rbt, \\
\frac{\mathrm{d}\rbt}{\mathrm{d}\tau} & = -\frac{\partial H(\ubt, \rbt)}{\partial\ubt} = -\frac{\partial U(\ubt)}{\partial\ubt}.
\label{eq:hamiltonsde}
\end{split}
\end{equation}
Algorithm~\ref{alg:hmc} shows the sampling with the HMC method.
\begin{algorithm}[!t]
\caption{HMC method sampling}
\label{alg:hmc}
{\footnotesize
\begin{algorithmic}[1]
\State Initialize $\ubt$ at random
\For{$l = 1, \dots, L_\text{HMC}$}
	\State Draw $\rbt$ from $\mathcal{N}(\zero, \Ibt)$
	\State \parbox[t]{1.0\linewidth}{Numerically solve Hamilton's equations~\eqref{eq:hamiltonsde} to obtain $\ubt^\prime$ and $\rbt^\prime$}
	\State \parbox[t]{1.0\linewidth}{Update $\ubt \leftarrow \ubt^\prime$ with probability min$[1,  \exp\{H(\ubt, \rbt) - H(\ubt^\prime, \rbt^\prime)\}]$}
	\State \parbox[t]{0.9\linewidth}{Regard the updated $\ubt$ as a sample from the posterior distribution $p(\ubt \mid \ybt)$}
\EndFor
\end{algorithmic}
}
\end{algorithm}
The Hamiltonian is constant based on \eqref{eq:Hamiltonian}.
Thus, a large change in the momentum $\rbt$ significantly influences the sample value $\ubt$. 
In addition, as per Algorithm~\ref{alg:hmc}, most proposals $\ubt^\prime$ are accepted with a probability of one, excluding cases of numerical errors. 
These lead to a typically higher sampling efficiency for the HMC method than for other MCMC methods.

According to Algorithm~\ref{alg:hmc}, when solving \eqref{eq:hamiltonsde} numerically during one step of the Markov chain, the log-posterior probability density derivative is internally evaluated $L$ times. The value of $L$ can vary depending on the problem and conditions, but it is typically assumed to be 10 in this study~\cite{BDA3}. The main factor contributing to the computational complexity is the calculation of the term $(\Hbt^\text{T}\Hbt)\ubt$. This term is included in the derivative of the log-likelihood contained in the log-posterior probability density, and because $\Hbt$ is assumed to be known, $(\Hbt^\text{T}\Hbt)$ only needs to be calculated once. Hence, the computational complexity of the HMC method is $\mathcal{O}(L N^2) = \mathcal{O}(10 N^2)$ per step of the Markov chain. With $L_\text{HMC}$ being the total steps in the Markov chain, the final computational complexity becomes $\mathcal{O}(10 L_\text{HMC} N^2)$.

\begin{figure*}[!t]
\centering
  \begin{minipage}[t]{0.338\linewidth}
    \centering
    \includegraphics[clip,page=1,width=\columnwidth]{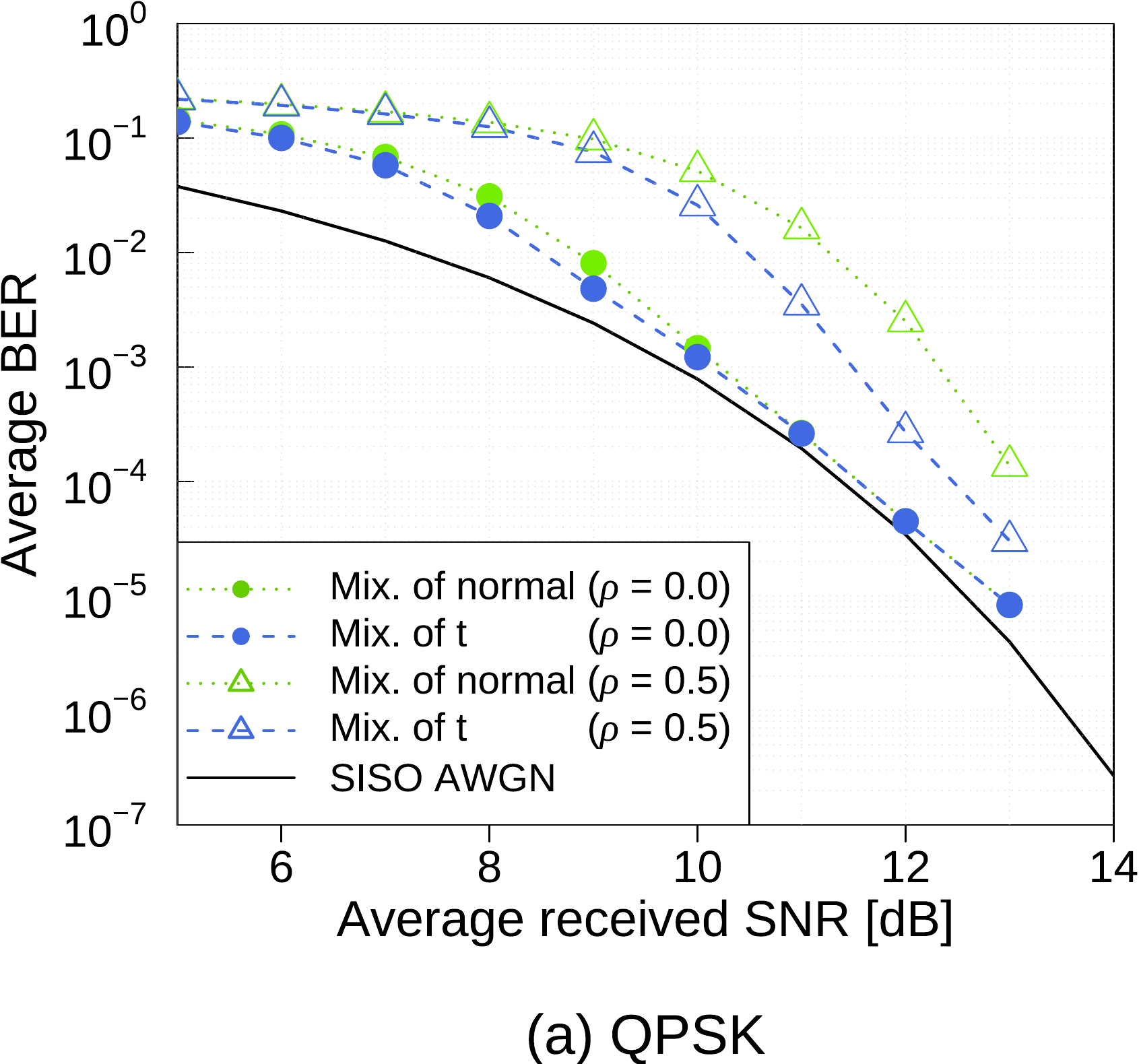}
  \end{minipage}
  \hspace{-0.01\linewidth}
  \begin{minipage}[t]{0.32\linewidth}
    \centering
    \includegraphics[clip,page=2,width=\columnwidth]{BER1-crop.pdf}
  \end{minipage}
 \hspace{-0.01\linewidth}
  \begin{minipage}[t]{0.32\linewidth}
    \centering
    \includegraphics[clip,page=3,width=\columnwidth]{BER1-crop.pdf}
  \end{minipage}
\caption{Average BER vs. average received SNR for $\rho = 0 \text{ and } 0.5$.}
\label{fig:tVSN}
\end{figure*}
%

\section{Numerical results and discussion} \label{sec:Numerical results and discussion}
We demonstrate that the proposed method can attain near-optimal signal detection in MIMO systems with polynomial computational complexity. 
To verify this, we perform a theoretical analysis of the computational complexity as well as computer simulations on signal detection.

\subsection{Settings in numerical analysis}
\paragraph*{Common assumptions}
We assume typical and exhaustive conditions, such as the number of antennas considering massive MIMO and the modulation order from low to high (as outlined in Table~\ref{tbl:simulationassumption}).
The BER plots are omitted if the computer simulation results are error-free.
\begin{table}[!t]
\caption{Common assumptions in numerical analysis}
\label{tbl:simulationassumption}
\centering
\begin{tabular}{cc} \hline
\noalign{\vspace{0.3mm}} 
\multicolumn{1}{c}{Item} & \multicolumn{1}{c}{Setting} \\ \hline
\noalign{\vspace{0.3mm}} 
Trials & 5000 \\
Number of antennas & $N = M = 96$ \\
Modulation order & QPSK, 16QAM, and 64QAM \\
Average transmission power & 1 \\
Fading & Quasi-static Rayleigh \\
Channel correlation & \parbox[t]{0.5\linewidth}{\linespread{0.8}\selectfont \centering Kronecker model \\(correlation coefficient  $\rho = 0$ or $0.5$)} \\
Channel coding& Uncoded \\ \hline
\end{tabular}
\end{table}

\paragraph*{Parameters of proposed method}
The parameters of the component's normal and $t$-\changecolor{distributions} are set to the values that displayed favorable performance in the preliminary rough search (Table~\ref{tbl:N_and_t_parameters}).
\begin{table}[!t]
\caption{Parameters of component's normal distribution and $t$-distribution}
\label{tbl:N_and_t_parameters}
\centering
\begin{tabular}{cccc} \hline
\noalign{\vspace{0.3mm}}
		&	Mixture of normal distributions		& \multicolumn{2}{c}{Mixture of $t$-distributions}	\\ \cline{3-4}
		&	$\sigma$						&	$\sigma$		 		& $\nu$ 				\\ \hline
\noalign{\vspace{0.3mm}}
QPSK	&	0.2483						&	0.5 $\times$ 0.2483	& 1.8 				\\
16QAM	&	0.1242						&	0.5 $\times$ 0.1242	& 1.8 				\\
64QAM	&	0.0664						&	0.8 $\times$ 0.0664	& 2.5 				\\ \hline
\end{tabular}
\end{table}
Each Markov chain in the simulation is assumed to have $2N$ steps, in line with the virtual parallelization approach of the MGS method. 
The parallel number of Markov chains is set to $\lfloor 1000/(2N) \rfloor$, which is the minimum number required to attain sufficient performance~\cite{WPMC}. 
As a result, the total number of steps in the Markov chain, $L_\text{HMC}$, is equivalent to 1000.

\paragraph*{Parameters of existing methods}
The total number of steps in the Markov chain of the MGS method is set to $L_\text{MGS} = 1000$, equal to $L_\text{HMC}$. 
However, the proposed method has a computational complexity that is 10 times greater than the MGS method due to its implicit internal loop $L = 10$. 
To ensure a fair comparison, 10 multiple restarts of the MGS method are performed. 

The total number of EP method iterations is set to $L_\text{EP} = 10$, which is a sufficient number to obtain adequate performance.

In the context of MIMO signal detection, the performance of a single-input single-output (SISO) transmission under additive white Gaussian noise (AWGN) serves as an ideal benchmark, as it eliminates inter-antenna interference and fading effects. The BER performance of the SISO AWGN is shown to highlight the theoretical lower bound of MIMO signal detection. 
\subsection{Computational complexity}
The computational complexities of the proposed method, the MGS method, and the EP method for signal detection are $\mathcal{O}(10 \times 1000 N^2)$, $10 \times \mathcal{O}(1000 N^2)$, and $\mathcal{O}(10 N^3)$, respectively, where $N = M$.
These complexities are all of polynomial orders.
Particularly, the proposed method and the MGS method have the same complexity under the given conditions, whereas the EP method is the least computationally expensive due to $N = 96\, (< 1000)$.

\subsection{BER performance: a mixture of $t$-distributions prior vs.~a mixture of normal distributions prior}
\paragraph*{Overall trend (Fig.~\ref{fig:tVSN} (a) through (c))}
The higher the modulation order, the further the performance deviates from the SISO AWGN.
This is because as the modulation order increases, the number of possible transmission signal points becomes larger, and the distance between these signal points becomes narrower, making the estimation more challenging.

\paragraph*{Comparison with a mixture of normal distributions prior (Fig.~\ref{fig:tVSN} (a) through (c))}
When $\rho = 0$, it is difficult to discern the difference in performance between the two methods as they overlap in certain areas. A clearer difference in performance can be observed when $\rho = 0.5$, which brings more challenging signal detection scenarios. For all modulation orders, the mixture of $t$-distributions prior is found to be superior when $\rho = 0.5$.
For instance, when the BER is $10^{-3}$, the mixture of $t$-distributions prior outperforms the mixture of normal distributions prior by 0.8, 0.5, and 0.3 dB for QPSK, 16QAM, and 64QAM, respectively.
In summary, the mixture of $t$-distributions prior provides better signal detection performance than the mixture of normal distributions prior due to its component's narrower peak and thicker tail. The former allows for a more focused search around the transmission signal points, whereas the latter enables a more aggressive search across the ``valley of death'' between the transmission signal points.

The improvement obtained from using a mixture of $t$-distributions as the prior distribution decreases as the modulation order increases. This advantage is particularly small for 64QAM modulation. 
As the modulation order increases and the solution space expands, even with an aggressive search using a $t$-distribution component, the effectiveness of the method becomes limited with a finite number of searches.

\subsection{BER performance: proposed method with a mixture of $t$-distributions prior vs.~existing methods}
\begin{figure*}[!t]
\centering
  \begin{minipage}[t]{0.33\linewidth}
    \centering
    \includegraphics[clip,page=1,width=\columnwidth]{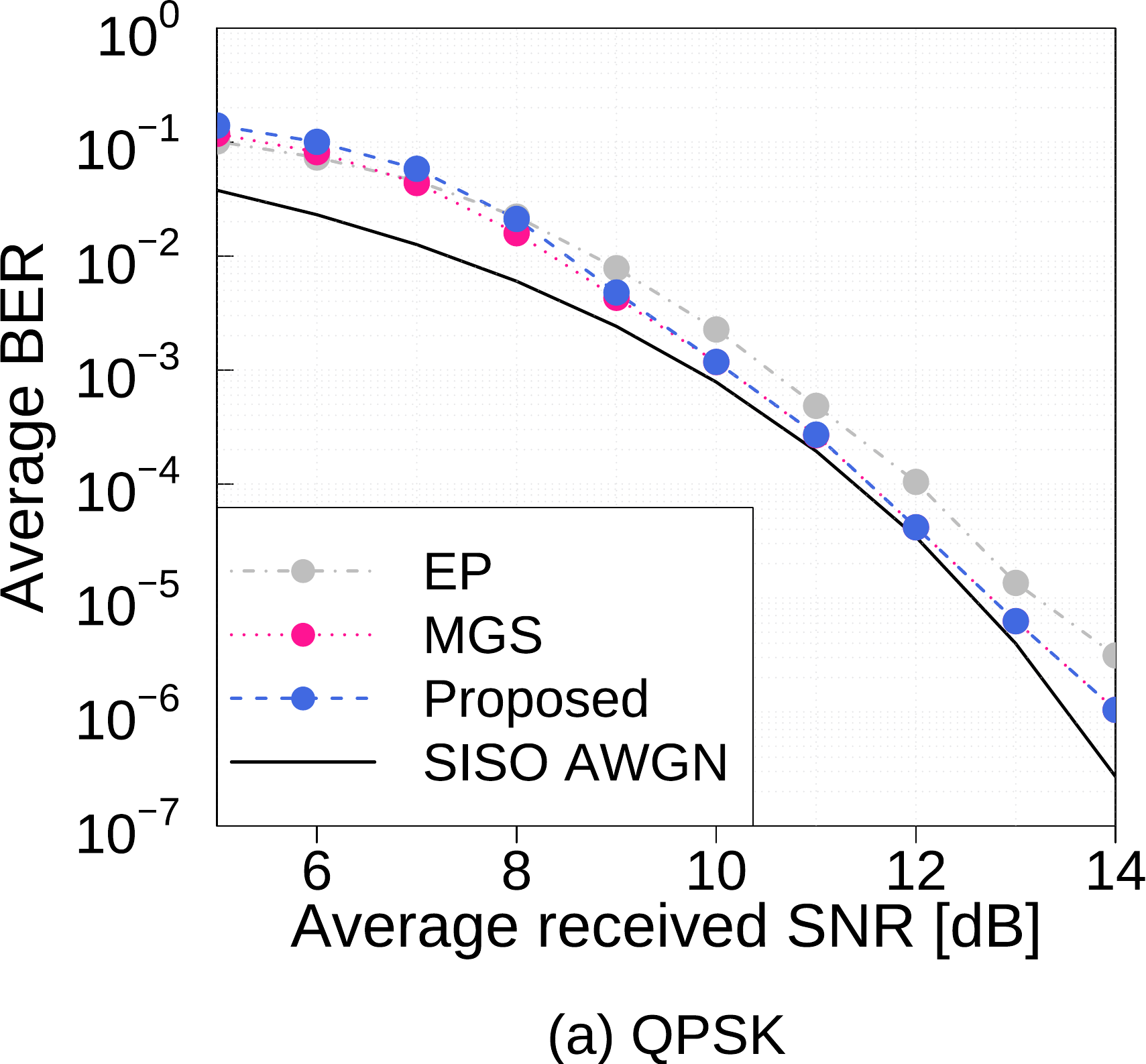}
  \end{minipage}
  \hspace{-0.01\linewidth}
  \begin{minipage}[t]{0.33\linewidth}
    \centering
    \includegraphics[clip,page=2,width=\columnwidth]{BER2-crop.pdf}
  \end{minipage}
 \hspace{-0.01\linewidth}
  \begin{minipage}[t]{0.33\linewidth}
    \centering
    \includegraphics[clip,page=3,width=\columnwidth]{BER2-crop.pdf}
  \end{minipage}
\caption{Average BER vs. average received SNR for $\rho = 0$.}
\label{fig:tVSE}
\end{figure*}

\paragraph*{Overall trend (Fig.~\ref{fig:tVSE} (a) through (c))}
As aforementioned, the higher the modulation order, the further the performance is away from the SISO AWGN.

\paragraph*{Comparison with existing methods for QPSK (Fig.~\ref{fig:tVSE} (a))}
The performance of all methods is comparable to the SISO AWGN, with only minute differences. This is due to the signal detection ease, given the limited number of potential transmission signal points and the significant distance between them.
In low signal-to-noise ratio (SNR) conditions, the proposed method performs slightly worse than the MGS and EP methods.
Under noisy conditions and limited searches, exploring areas outside the transmission signal points appears less effective. In such cases, the original impulse-like prior distribution used in the MGS and EP methods appears to be sufficient.
It is noted that a low SNR region in practical communication environments usually requires the combination of other improvement techniques, such as retransmission protocol and error-correcting code.
Consequently, this slight inferiority of the proposed method is expected to have only a limited impact on the overall system performance.
Meanwhile, at moderate-to-high SNRs, the proposed method outperforms the EP method and is comparable to the MGS method. For instance, the SNR gain of the proposed method at a BER of $10^{-3}$ is close to 0 and 0.4 dB compared with the MGS and EP methods, respectively.

\paragraph*{Comparison with existing methods for 16QAM (Fig.~\ref{fig:tVSE} (b))}
Compared with the QPSK case, the 16QAM has an increased number of possible transmission signal points and a narrower distance between these points, making signal detection more challenging.
Such a difficult condition makes the characteristics of each method more clear.
The difference from the QPSK case is that the proposed method outperforms the MGS method in terms of detection performance at moderate-to-high SNRs, resulting in the best detection performance for the proposed method at moderate-to-high SNRs.
In addition, compared with the QPSK case, the performance difference between the proposed method and existing methods is larger at moderate-to-high SNRs.
For instance, the SNR gain of the proposed method at a BER of $10^{-3}$ is close to 0.4 and 1.6 dB compared with the MGS and EP methods, respectively.

\paragraph*{Comparison with existing methods for 64QAM (Fig.~\ref{fig:tVSE} (c))} 
Compared with the QPSK and 16QAM cases, signal detection for 64QAM is the most challenging, and the distinctiveness of the proposed method is the most pronounced. This is due to the greatest number of potential transmission signal points and the smallest distance between these points. The difference between this case and the QPSK and 16QAM cases is that the proposed method outperforms the MGS method in terms of detection performance at low SNR. Although the proposed method demonstrates its best performance at moderate-to-high SNRs, similar to the 16QAM case, the difference in performance with existing methods is greater than that seen in the 16QAM case. 
For instance, at a $10^{-3}$ BER, the proposed method shows an indeterminable large gain in SNR compared with the MGS method and a 4.2-dB gain compared with the EP method.

\changecolor{%
Additional verification is performed here because the BER performance of MGS is extremely degraded compared to QPSK and 16QAM.
Fig.~\ref{fig:MGSreconsider} shows excerpts from Figs. 5 (c) and 6 (c) in \cite{WPMC}.
The ``Conv.'' and the ``iterations'' in the legend indicate the MGS and the total number of iterations of the Markov chain, respectively.
Fig.~\ref{fig:MGSreconsider} shows that BER performance of MGS improves with fewer transmitting and receiving antennas and a higher number of iterations of the Markov chain.
Compared to \cite{WPMC}, the assumptions of this study make signal detection more difficult due to more transmitting and receiving antennas and fewer iterations of the Markov chain.
Therefore, the result of MGS is considered reasonable.
\begin{figure*}[!t]
\centering
\includegraphics[clip,page=1,width=1.0\linewidth]{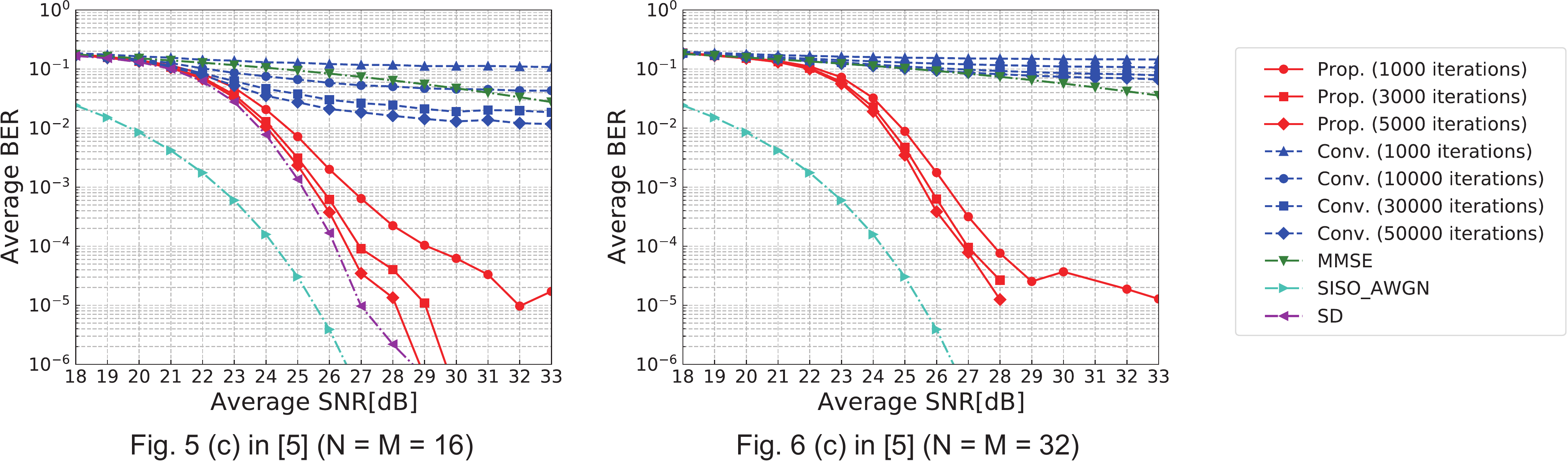}
\caption{\changecolor{Average BER vs. average received SNR for 64QAM and $\rho = 0$. Reprinted from~\cite{WPMC} with permission (\copyright\,2021 IEEE).}}
\label{fig:MGSreconsider}
\end{figure*}
}
\subsection{BER performance: proposed method with a mixture of $t$-distributions prior vs.~SISO AWGN (Fig.~\ref{fig:tVSE} (a) through (c))} 
The proposed method achieves near-optimal performance. Specifically, the SNR degradation of the proposed method at a $10^{-3}$ BER is within 0.3, 1.3, and 2.9 dB for QPSK, 16QAM, and 64QAM, respectively. This exceptional performance is achieved through a combination of the appropriate prior distribution setting and an effective search algorithm. In particular, the proposed prior distribution is sufficiently similar to the discrete point mass at the transmission signal point and also enables active exploration of other potential signal points. Therefore, we consider the proposed method is sufficiently reliable, as demonstrated by its outstanding performance and appropriate methodology.

\section{Conclusions} \label{sec:Conclusions}
Based on a mixture of $t$-distributions prior and the HMC method, we proposed a signal detection method that is expected to deliver near-optimal performance with polynomial computational complexity. In terms of detection performance, the proposed method showed a substantial improvement over typical existing methods for higher-order modulation. Higher-order modulation could be crucial for future wireless communications. The amount of data transferred in wireless communications is projected to grow exponentially with expectations for multi-sensory interactions beyond traditional voice and video communication. 
In this sense, the proposed method is deemed beneficial.

The limitations of this study are as follows. The proposed method's computational complexity, although of polynomial order, is greater than that of the EP method. In addition, the BER performance may be slightly lower than those of the MGS and EP methods at low SNR. Fine-tuning the $t$-distribution parameter for improved performance at low SNR may be possible, but further investigation is required.
\changecolor{Although channel coding was not applied in this study, signal detection methods are often used in combination with error-correcting codes in practice.
Therefore, we will carefully investigate its impact in the future, considering various parameters such as encoding scheme, code rate, decoding algorithm, and interleaving.}

From a mathematical perspective, our study suggests that extending the discrete problem to a continuous one can lead to better solutions. This approach has the potential to revolutionize combinational optimization, including well-known problems like the traveling salesman and nurse scheduling problems.


\end{document}